# Universality away from critical points in two-dimensional phase transitions

Cintia M. Lapilli, Peter Pfeifer, and Carlos Wexler

*Department of Physics and Astronomy, University of Missouri, Columbia, Missouri, 65211, USA*

(March 8, 2006)

The $p$-state clock model in two dimensions is a system of discrete rotors with a quasi-liquid phase in a region $T_1 < T < T_2$ for $p > 4$. We show that, for $p > 4$ and above a temperature $T_{eu}$, all macroscopic thermal averages, such as energy or magnetization, become identical to those of the continuous rotor ($p = \infty$). This *collapse of thermodynamic observables* creates a regime of *extended universality* in the phase diagram and an *emergent symmetry*, not present in the Hamiltonian. For $p \geq 8$, the collapse starts in the quasi-liquid phase and makes the transition at $T_2$ identical to the Berezinskii-Kosterlitz-Thouless (BKT) transition of the continuous rotor. For $p \leq 6$, the transition at $T_2$ is below $T_{eu}$ and no longer BKT. The results generate a comprehensive map of the critical properties at $T_1$ and $T_2$, and a range of experimental predictions, such as motion of magnetic domain walls, fabrication of identical devices from different building blocks, and limits on macroscopic distinguishability of different microscopic interactions.

A cornerstone in the study of phase transitions is the concept of universality, stating that entire families of systems behave identically in the neighborhood of a critical point, such as the liquid-gas critical point in a fluid or the Curie point in a ferromagnet, at which two phases become indistinguishable. Near the critical point, thermodynamic observables do not depend on the details of intermolecular interactions, and the critical exponents, which quantify how observables go to zero or infinity at the transition, depend only on the range of interactions, symmetries of the Hamiltonian, and spatial dimensionality of the system. Universality arises as the system develops fluctuations of all sizes near the critical point, which wash out the details of interaction and render the system scale-invariant [1,2].

Here we report a new, stronger form of universality. We find the remarkable result that, in a specific family of systems, different members behave identically both near and away from critical points—we call this *extended universality*—if the temperature exceeds a certain value $T_{eu}$. In this regime, universality occurs not just at a critical point but over a whole range of temperatures; yields identical values of *all* (macroscopic) *thermodynamic observables* (such as energy or magnetization), not just identical critical exponents—we call this *collapse of thermodynamic observables*—for different systems; runs over Hamiltonians with different symmetries; and is not induced by large fluctuations. As the collapse maps Hamiltonians with different symmetries onto one and the same thermodynamic state, the system exhibits a symmetry not present in the Hamiltonian. The added symmetry at high temperature is the counterpart of broken symmetry at low temperature. To the best of our knowledge, no such collapse of thermodynamic observables has been observed before.

The family under consideration is the $p$-state clock model, also known as $p$-state vector Potts model or $Z_p$ model [3], in two dimensions, with Hamiltonian

$$H_p = -J_0 \sum_{\langle i,j \rangle} \mathbf{s}_i \cdot \mathbf{s}_j = -J_0 \sum_{\langle i,j \rangle} \cos(\theta_i - \theta_j), \quad (1)$$

where each dipole or spin, $\mathbf{s}_i$, can make $p$ angles $\theta_i = 2\pi n_i/p$ ($n_i = 1, ..., p$), the sum is over nearest neighbors on a square lattice, and the coupling is ferromagnetic, $J_0 > 0$. The $p$ discrete orientations, imposed by a crystallographic substrate and molecular shapes, make $H_p$ invariant under the transformations $\theta_i \rightarrow \theta_i + 2\pi n/p$, $n = 1, ..., p$ (cyclic group $Z_p$). The model interpolates between spin up/down of the Ising model [4] ($p = 2$) and the continuum of directions of the planar rotor, or XY, model [5,6] ($p = \infty$). It has been studied to track how the phase transition in the Ising model, with spontaneously broken symmetry in the ferromagnetic phase, gives way to the BKT transition [6], without broken symmetry, in the rotor model. As the symmetry of $H_p$ changes with $p$, so may the universality class of the phase transitions.

Elitzur *et al.* [7] showed that the model has a rich phase diagram: for $p = 2, 4$, it belongs to the Ising universality class, with a low-temperature ferromagnetic phase and a high-temperature paramagnetic phase; for $4 < p < \infty$, three phases exist—a low-temperature ordered and a high-temperature disordered phase, as in the Ising model, plus a quasi-liquid intermediate phase. Duality transformations [7,8] and renormalization group (RG) treatments [9,10] shed light on the phases in terms of a related model,

$$H_{\{h_p\}} = -J_0 \sum_{\langle i,j \rangle} \cos(\theta_i - \theta_j) + \sum_i \sum_p h_p \cos(p\theta_i), \quad (2)$$

where the $\theta_i$'s are continuous and the $h_p$'s are symmetry-breaking fields, mimicking the constraint to $p$ spin directions in the clock model. The clock model obtains by letting $h_p \rightarrow \infty$ for a selected $p$. José *et al.* [10] showed, in a self-dual approximation of (2), that the fields were relevant for $p < 4$, and irrelevant for $p > 4$. But (1) is not self-dual for $p > 4$, and RG approximations examining the effect of the discreteness of the angular variables are delicate near $p = 6$. As a result, the transition points of (1) in the three-phase region are not precisely known.

We establish the phase diagram, collapse of observables, and associated temperature $T_{eu}$ by Monte Carlo

(MC) simulations [11,12]. The simulations were performed on a square lattice of size $N = L \times L$, with $L = 8$–72, and periodic boundary conditions. Averages involved sampling of $10^5$–$10^5$ configurations, with equilibration runs of $p \times (1{,}000$–$5{,}000)$ MC steps, a step being one attempt to change *every* spin.

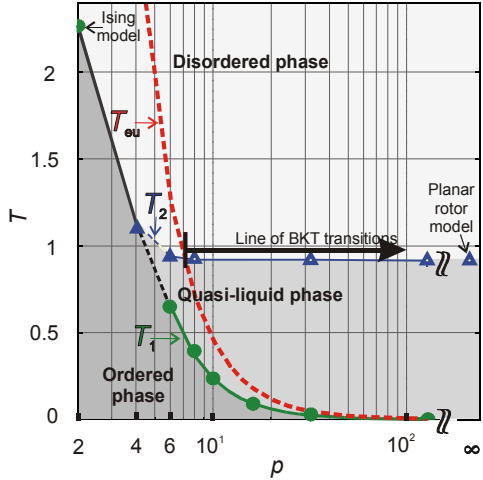

FIG. 1 (color online). Phase diagram of the *p*-state clock model. The Ising model, $p = 2$, exhibits a single second-order phase transition, as does the $p = 4$ case, which is also in the Ising universality class. For $p > 4$, a quasi-liquid phase appears, and the transitions at $T_1$ and $T_2$ are both second-order. The line $T_{\text{eu}}$ separates the phase diagram into a region where the thermodynamic observables do depend on $p$, below $T_{\text{eu}}$, and a region where their values are *p*-independent, above $T_{\text{eu}}$ (collapse of observables, extended universality). For $p \geq 8$, we observe $T_{\text{eu}} < T_2 = T_{\text{BKT}} \simeq 0.89$. Throughout, temperatures are in units of $J_0/k_B$, where $k_B$ is Boltzmann's constant, and energies are in units of $J_0$.

Figure 1 summarizes our results. The Ising model shows the expected phase transition at $T_c^{\text{Ising}} = 2/\ln[1+\sqrt{2}] \simeq 2.27$. The case $p = 4$ also shows a single transition, at $T_c = T_c^{\text{Ising}}/2 \simeq 1.13$. Most interesting is the case $p > 4$. It hosts the two transitions predicted by Elitzur *et al.* [7], illustrated in Fig. 2, and the collapse of thermodynamic observables at $T > T_{\text{eu}}$. At $T_{\text{eu}}$ the system switches from a *p*-dependent state ($T < T_{\text{eu}}$; discrete symmetry) to a state indistinguishable from $p = \infty$ ($T > T_{\text{eu}}$; continuous symmetry). For $p \leq 4$ there is no collapse.

We characterize the transitions using Binder's fourth-order cumulants [11] in magnetization, $U_L \equiv 1 - \frac{1}{3}\langle m^4 \rangle / \langle m^2 \rangle^2$, and energy, $V_L \equiv 1 - \frac{1}{3}\langle e^4 \rangle / \langle e^2 \rangle^2$. The high-temperature transition, $T_2$, is obtained from the fixed point of $U_L$. The latent heat, proportional to $\lim_{L\to\infty} [2/3 - \min_T V_L]$, vanishes, signaling a second-order transition. The low-temperature transition, $T_1$, is obtained from the temperature derivative of the magnetization, $\partial \langle |m| \rangle / \partial T$, and $\partial U_L / \partial T$, which diverge as $L \to \infty$ (thermodynamic limit). Finite-size scaling (FSS) of the derivatives' minima yields $T_1 = \lim_{L\to\infty} T_{1,L}$. We find $T_1 = 4\pi^2/(\tilde{T}_2 p^2)$, with $\tilde{T}_2 \simeq 1.67 \pm 0.02$; the ordered phase vanishes rapidly as $p \to \infty$.

Figure 3 shows selected thermodynamic observables: heat capacity, $c_F \equiv (\langle H^2 \rangle - \langle H \rangle^2)/[L^2 T^2]$, and magnetization, $\langle \mathbf{m} \rangle \equiv \langle \mathbf{M} \rangle / L^2 = \langle (|\sum_{i=1}^N \cos\theta_i|, |\sum_{i=1}^N \sin\theta_i|) \rangle / L^2$, per spin. The Ising-like behavior for $p = 4$, and the three phases for $p > 4$ are evident. Figure 3 proves the collapse of thermodynamic observables (in the thermodynamic limit): $c_F$ and $\langle |\mathbf{m}| \rangle$ are manifestly *p*-independent for $p > 4$ and $T > T_{\text{eu}}$.

$$T_{\text{eu}} = \frac{4\pi^2}{p^2 T_{\text{BKT}}}, \qquad (3)$$

where $T_{\text{BKT}} \simeq 0.89$; and the internal-energy differences abruptly vanish at $T = T_{\text{eu}}$.

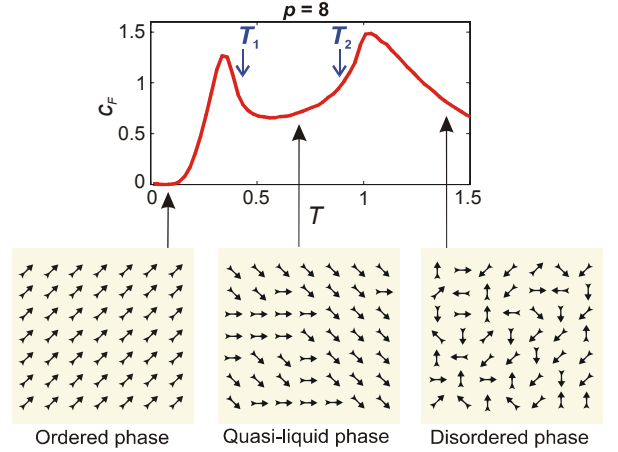

FIG. 2 (color online). Three-phase regime for $p = 8$, in terms of the specific heat, $c_F$, transition temperatures $T_1$, $T_2$ (the transitions do *not* occur at the peaks of $c_F$), and typical spin configurations. The correlation function, $\langle \mathbf{s}_i \cdot \mathbf{s}_j \rangle$, goes to a non-zero constant at large distance $|i - j|$, at $T < T_1$ (long-range order), decays as a power law of distance at $T_1 < T < T_2$ (quasi-long-range order [5,6]; typical configurations contain vortices, such as the clockwise vortex fragment in the center panel), and decays exponentially with distance at $T > T_2$ (disorder).

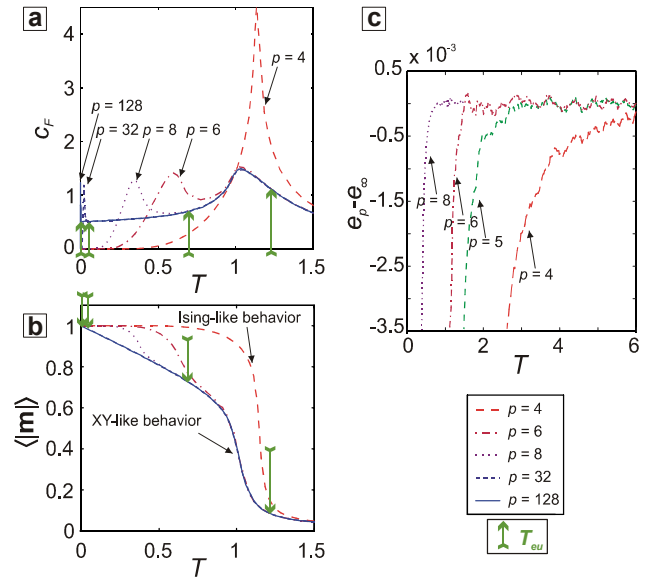

FIG. 3 (color online). Heat capacity (a), magnetization (b), and difference of internal energy per spin relative to the planar rotor model (c). The data correspond to a system size of $L = 72$ ($N = 5{,}184$ spins). All curves coalesce above $T_{\text{eu}}$ (arrows) for $p \geq 5$ (collapse of thermodynamic observables, extended universality).

The specific form of $T_{eu}(p)$ can be understood as follows. (i) The large-$p$, small-$(\theta_i - \theta_j)$ expansion of (1) yields a characteristic temperature, $\sim (2\pi/p)^2$, such that all averages become $p$-independent whenever $T/(2\pi/p)^2 \gg 1$, implying an *asymptotic* collapse of observables. (ii) Elitzur et al. [7] noted that discreteness of the angles $\theta_i$ becomes irrelevant for the critical properties of (1), for sufficiently large $p$, implying the collapse of observables at *critical points*, $T_2$. (iii) A similar irrelevance of the discreteness of angles, imposed by $h_p \to \infty$, was observed for (2), subject to $T > 4\pi^2/(p^2 T_k)$, where $T_k \simeq 1.35$ is the BKT point of the self-dual approximation of (2) [10]. (iv) For the *full* collapse of thermodynamic observables in the clock model, these partial results suggest that a necessary condition for collapse is $T > 4\pi^2/(p^2 T_{BKT})$. The fit of our data for $T_{eu}(p)$, yielding (3), validates this expectation and shows that the condition is necessary and sufficient.

The collapse/non-collapse above/below the curve $T_{eu}(p)$ makes far-reaching predictions for the transitions $T_1$ and $T_2$, which we now test. We begin with $T_2$. We observe that $T_2 > T_{eu}$ for $p \geq 8$, which implies that the transition $T_2$ *must* be BKT for $p \geq 8$. Previous work advanced only the plausibility of such universality. To test our assertion, beyond the equality $T_2 = T_{BKT}$, we equate BKT behavior to the following planar-rotor properties [6]: (i) discontinuous jump to zero of the helicity modulus, $\Upsilon(T_{BKT}^-) = 2T_{BKT}/\pi$; (ii) exponentially diverging correlation length, $\xi \sim \exp[c/|T - T_{BKT}|^{1/2}]$; (iii) temperature-dependent power-law decay of two-point correlation functions, with exponent $\eta(T_{BKT}) = 1/4$; (iv) decay of the magnetization, with exponent $\beta = 3\pi^2/128$ [13].

Our simulations fully confirm these properties at $T_2$ and $p \geq 8$ [12]. We illustrate this for the discontinuity of the helicity modulus. Following the Minnhagen-Kim stability argument [14], we evaluate the change in free energy when a twist $\Delta$ is applied to the spins:

$$f = \frac{\Upsilon}{2}\Delta^2 + \frac{\Upsilon_4}{4!}\Delta^4 + \cdots. \quad (4)$$

Figure 4 shows $\Upsilon$ and $\Upsilon_4$ (fourth-order helicity) as a function of $T$ and system size $L$ for $p = 8$. At $T_2$, $\lim \Upsilon_4 < 0$, so if $\lim_{L\to\infty} \Upsilon$ went to zero continuously as $T \to T_2^-$, the free energy would turn negative and the system would become unstable as $T \to T_2$. This contradiction implies that $\Upsilon$ goes to zero discontinuously. The same result is obtained for all $p \geq 8$, as predicted by the collapse of observables. Conversely, the non-collapse of observables at $T_2$ for $p \leq 7$ suggests that the transition at $p \leq 7$ differs from BKT. This is indeed the case: $\Upsilon$ does not vanish, and $\Upsilon_4$ converges to zero as $L \to \infty$. The nonzero helicity modulus and its continuity at $T_2$ make the transition manifestly non-BKT, according to our criterion. Other critical properties computed at $T_2$ also differ from the BKT values [12]. Moreover, visibly $T_2 > T_{BKT}$ (Fig. 1). Thus, contrary to prior conjectures that the transition at $T_2$ and $p = 6, 7$ is BKT-like [7,10,15,16], we find that it differs significantly from BKT. Specifically, our analysis shows that a twist at $T_2^+$, costs much more energy, $f = \frac{1}{2}\Upsilon\Delta^2 + O(\Delta^6)$, than in the BKT case, $f = O(\Delta^6)$.

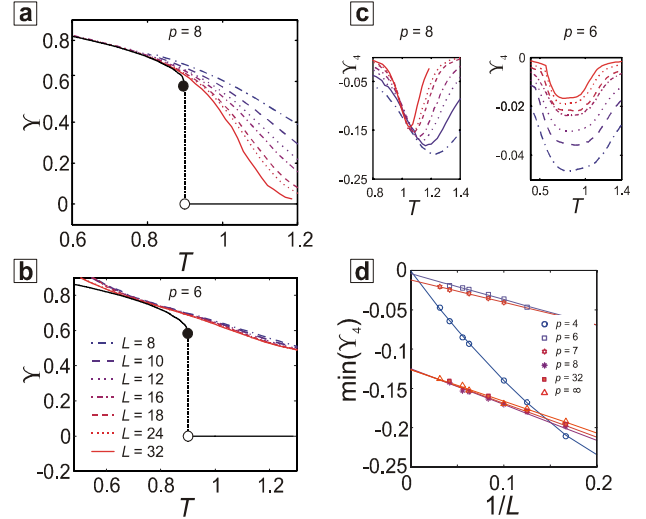

FIG. 4 (color online). Helicity modulus, $\Upsilon$, and fourth-order helicity, $\Upsilon_4$, for $p = 8$ (a, c) and $p = 6$ (b, c) across the phase transition $T_2$. The bottom curve in (a, b), for reference, is the modulus for the planar rotor, which jumps from $2T_{BKT}/\pi$ (full circle) to zero at $T = T_{BKT}$. For all $p \geq 8$, $\lim_{L\to\infty} \Upsilon = 0$. Extrapolation of $\Upsilon_4$ to the thermodynamic limit yields two classes of results (d): $\Upsilon_4$ converges to the universal value $-0.126 \pm 0.005$ for $p \geq 8$, and to zero for $p \leq 7$. This implies that $\Upsilon$ has a discontinuous jump to zero at $T_2$ if and only $p \geq 8$ (see text).

We turn to the low-temperature transition, $T_1$, which also has been argued to be BKT-like for $p \geq 6$ [16,17]. The non-collapse of observables at $T_1$ for all finite $p$, suggests, and our simulations substantiate [12], that this transition, too, differs significantly from BKT. E.g., our FSS analysis of the temperature derivatives of the magnetization and its fourth-order cumulant gives a power-law dependence $T_{1,L} = T_1 + O(1/L)$ [12], as expected when the low-temperature phase exhibits long-range order. In contrast, BKT would give $T_{1,L} = T_1 + O[(\ln L)^{-2}]$ [13,16,17].

Thus the collapse of thermodynamic observables has remarkable consequences on the phase diagram of the clock model and resolves longstanding questions of similarities and differences with the planar rotor model. When present, $T > T_{eu}$, the collapse causes the spins to lose their identity as discrete-symmetry variables and become indistinguishable from the continuous-symmetry variables of the planar rotor. At critical points, $T_2$ for $p \geq 8$, it guarantees that all critical properties are identical to those of the BKT transition. Away from critical points, it guarantees that the quasi-liquid phase and disordered phase are identical to those of the planar rotor. When the collapse is absent, $T < T_{eu}$, the spins retain their discrete symmetry, and all critical points, $T_2$ for $p < 8$ and $T_1$ for $p < \infty$, are distinctly non-BKT.

Our results raise important questions and implications. Just as universality at a critical point is accompanied by invariance of the system under the scale transformation $\mathbf{r}_i \to \lambda \mathbf{r}_i$, even though the Hamiltonian has no such symmetry, extended universality is accompanied by invariance under the transformation $\theta_i \to \theta_i + \alpha$, $\alpha$ arbitrary, even though (1) is invariant only under discrete rotations. Thus both universalities are generated by an *emergent symmetry*, not present in the Hamiltonian. What makes extended universality different is that the symmetry is present over a whole range of temperatures, not just at the critical point, giving it the status of a "protected" property [18] over a correspondingly wide range of temperatures. This suggests that thermal averages at $T > T_{eu}$ should be expressible in terms of a coarse-grained Hamiltonian invariant under rotation by $\alpha$, and that this representation is exact. The construction of such a representation, and accordingly the *origin* of extended universality, is an open problem. Related questions are: If the thermodynamic observables show $p$-dependent ferromagnetic ordering below $T_1$, but all $p$-dependence is lost above $T_{eu}$, what is the nature of the region $T_1 < T < T_{eu}$? Is the transition from uncollapsed to collapsed at $T_{eu}$, at fixed $p$, a phase transition in itself? If so, what is the nature of the nonanalyticity at $T_{eu}$?

A range of experimental systems, from thin magnetic films to monolayers of adsorbed molecules [19], have been modeled by dipoles restricted to $p$ orientations. Our results imply that $p$-state characteristics can be observed only at low temperatures, $T < T_{eu}(p)$. On the high side, $T > T_{eu}(p)$, we expect the results to be relevant for the design of dense layers of supercritically adsorbed gases for fuel storage [20,21], and vortex dynamics to move magnetic domain walls in "magnetic race-track memory" [22]. E.g., the low energy cost of a twist for $p \geq 8$ suggests an easy motion of domain walls. The collapse of observables may be studied directly in a monolayer of rotaxane, a molecular wheel threaded by a molecular axle, on a single-crystal surface [23]. If the wheels have $p$-fold symmetry, modulo a polar group mediating the interaction between neighboring wheels, their dynamics should be governed by (1). For $p = 8$, the hallmark of the collapse will be that the heat capacity peaks at $T \simeq 0.37$ and coalesces with the $p = \infty$ curve at $T = T_{eu} \simeq 0.69$. The collapse may also induce a change in the NMR signal of the polar group, as the group switches from a discrete rotor at $T < T_{eu}$ to a continuous rotor at $T > T_{eu}$. Other experimental platforms may be rotors with a magnetic ion at the center, driven by light [24].

It is a tenet in statistical mechanics that there is a one-to-one correspondence between microscopic dynamics and macroscopic observations. Our results present a significant counterexample: systems with different Hamiltonians may produce identical thermodynamics, over a wide range of temperatures. How ubiquitous is this phenomenon? If common, it would offer opportunities to fabricate equivalent devices, such as magnetic race tracks, from inequivalent building blocks. But the many-to-one map of Hamiltonians onto thermodynamic states equally demonstrates previously unknown limits on macroscopic distinguishablity of different microscopic interactions and raises the question, how can such interactions be distinguished experimentally?

We thank H. Fertig, G. Vignale, H. Taub, and K. Knorr for useful discussions. Acknowledgment is made to the University of Missouri Research Board and Council, the Donors of the Petroleum Research Fund, administered by the American Chemical Society, and the National Science Foundation (Grant No. EEC-0438469), for support of this research.

---

1. L.D. Landau, and E.M. Lifshitz, *Statistical Physics* (MIT Press, London, 1966).
2. L.P. Kadanoff *et al.*, *Rev. Mod. Phys.* **39**, 395 (1967); A.A. Migdal, *Sov. Phys. JETP* **42**, 743 (1976); L.P. Kadanoff, in *Proceedings of 1970 Varenna Summer School on Critical Phenomena*, ed. M.S. Green (Academic Press, New York, 1970); E. Wigner, *Physics Today* March 1964, p. 34.
3. R. Potts, *Proc. Camb. Phil. Soc.* **48**, 106 (1952).
4. L. Onsager, *Phys. Rev. B* **65**, 117 (1944).
5. N.D. Mermin, and H. Wagner, *Phys. Rev. Lett.* **17**, 1133 (1966).
6. V.L. Berezinsky, *Sov. Phys. JETP* **32**, 493 (1970); J.M. Kosterlitz, and D.J. Thouless, *J. Phys. C* **6**, 1181 (1973); J.M. Kosterlitz, *J. Phys. C* **7**, 1046 (1974); D.R. Nelson, and J.M. Kosterlitz, *Phys. Rev. Lett.* **39**, 1201 (1977).
7. S. Elitzur, R.B. Pearson, and J. Shigemitsu, *Phys. Rev. D* **19**, 3698 (1979).
8. R. Savit, *Rev. Mod. Phys.* **52**, 453 (1980).
9. J. Villain, *J. Physique* **36**, 581 (1975).
10. J.V. José, L.P. Kadanoff, S. Kirkpatrick, and D.R. Nelson, *Phys. Rev. B* **16**, 1217 (1977).
11. K. Binder, and D.W. Heermann, *Monte Carlo Simulation in Statistical Physics* (Springer, Berlin, 4th ed., 2002).
12. Details will be published elsewhere.
13. S.T. Bramwell, and P.C.W. Holdsworth, *J. Phys.: Condens. Matter* **5**, L53 (1993).
14. P. Minnhagen, and B.J. Kim, *Phys. Rev. B* **67**, 172509 (2003).
15. M.S.S. Challa, and D.P. Landau, *Phys. Rev. B* **33**, 437 (1986).
16. Y. Tomita, and Y. Okabe, *Phys. Rev. Lett.* **86**, 572 (2001).
17. E. Rastelli, S. Regina, and A. Tassi, *Phys. Rev. B* **69**, 174407 (2004).
18. R.B. Laughlin *et al*, *Proc. Nat. Acad. Sci.* USA **97**, 32 (2000).
19. E.g., S. Faßbender, M. Enderle, K. Knorr, J.D. Noh, and H. Rieger, *Phys. Rev. B* **65**, 165411 (2002).
20. P. Pfeifer *et al.*, *Phys. Rev. Lett.* **88**, 115502 (2002).
21. S. Patchkovskii, *et al.*, *Proc. Nat. Acad. Sci.* USA **102**, 104439 (2005).
22. S.S.P. Parkin, U.S. Patents 6,834,005 (2004), 6,898,132 (2005), and 6,920,062 (2005).
23. E.g., Y.H. Jang, S.S. Jang, and W.A. Goddard, III, *J. Am. Chem. Soc.* **127**, 4959 (2005).
24. M.F. Hawthorne *et al.*, *Science* **303**, 1849 (2004).